\documentclass[%
superscriptaddress,
reprint,
amsmath,amssymb,
aps,
prx,
10pt,
twocolumn,
floats,
]{revtex4-2}

\usepackage[english]{babel}
\usepackage[utf8]{inputenc}
\usepackage[T1]{fontenc}
\usepackage{graphicx}
\usepackage[caption=false]{subfig}
\usepackage{float}
\usepackage{dcolumn}
\usepackage{blindtext}
\usepackage{upgreek}
\usepackage{siunitx}
\usepackage{xcolor}

\DeclareSIUnit\gauss{G}
\usepackage{braket} 

\newcommand{\kcirc}[1]{\ket{#1 \mathrm{C}}}

\newcommand{\vv}[1]{\boldsymbol{#1}}
\newcommand{\e}{\mathrm{e}}
\newcommand{\deriv}[2]{\frac{\mathrm{d} #1}{\mathrm{d} #2}}
\newcommand{\dderiv}[2]{\frac{\mathrm{d}^2 #1}{\mathrm{d} #2^2}}

\newcommand{\norm}[1]{\Vert #1 \rVert}

\begin{document}

\title{Interacting Circular Rydberg Atoms Trapped in Optical Tweezers}
\author{P.~M\'ehaignerie}
\altaffiliation{These authors contributed equally to this work.}
\author{Y.~Machu}
\altaffiliation{These authors contributed equally to this work.}
\author{A.~Durán~Hernández}
\author{G.~Creutzer}
\affiliation{Laboratoire Kastler Brossel, Coll\`ege de France, CNRS, ENS-Universit\'e PSL, Sorbonne Universit\'e, 11 place Marcelin Berthelot, F-75231 Paris, France}
\author{D.~J.~Papoular}
\affiliation{LPTM, UMR 8089 CNRS \& CY Cergy Paris Université, Cergy-Pontoise, France}
\author{J.~M.~Raimond}
\affiliation{Laboratoire Kastler Brossel, Coll\`ege de France, CNRS, ENS-Universit\'e PSL, Sorbonne Universit\'e, 11 place Marcelin Berthelot, F-75231 Paris, France}
\author{C.~Sayrin}
\email[Corresponding author: ]{clement.sayrin@lkb.ens.fr}
\affiliation{Laboratoire Kastler Brossel, Coll\`ege de France, CNRS, ENS-Universit\'e PSL, Sorbonne Universit\'e, 11 place Marcelin Berthelot, F-75231 Paris, France}
\affiliation{Institut Universitaire de France, 1 rue Descartes, 75231 Paris Cedex 05, France}
\author{M.~Brune} 
\affiliation{Laboratoire Kastler Brossel, Coll\`ege de France, CNRS, ENS-Universit\'e PSL, Sorbonne Universit\'e, 11 place Marcelin Berthelot, F-75231 Paris, France}

\date{\today}

    \begin{abstract}
Circular Rydberg atoms (CRAs), i.e., Rydberg atoms with maximal orbital momentum, ideally combine long coherence times and strong interactions, a key property of quantum systems, in particular for the development of quantum technologies. However, the dipole-dipole interaction between CRAs has not been observed so far. 
We report the measurement and characterization of the resonant dipole-dipole interaction between two CRAs, individually trapped in optical tweezers, and find excellent agreement with theoretical predictions. We demonstrate a dynamic control over the strength of the interaction by tuning the orientation of an electric field. We use the interaction between the CRAs as a meter for the interatomic distance, and record the relative motion between two atoms in their traps. This motion, that we induce through the interaction between Rydberg levels with permanent electric dipoles, transiently populated during the preparation of the circular states, is a signature of spin-motion coupling.
    \end{abstract}

\maketitle    
The manipulation of individual quantum systems, whether in the form of atoms~\cite{Gross2017, Adams2019, Henriet2020, Amico2022}, ions~\cite{Monroe2021}, molecules~\cite{Kaufman2021, Langen2024, Deiss2024}, superconducting qubits~\cite{Devoret2013, Heras2014, Salathe2015, Arute2019, Wu2021}, quantum dots or single photons~\cite{Hartmann2016, Flamini2018, Romero2024} has tremendously progressed in the last decades, with the advent of quantum technologies relying on their mutual interactions or coupling to external fields~\cite{Altman2021, Oh2024}.  The performance of these quantum devices is limited by decoherence, making it essential to maximize the ratio between decoherence and characteristic interaction times. 

Circular Rydberg atoms (CRAs) are particularly interesting in this context. These giant states, with high principal quantum number, $n$, and maximal orbital momentum, $\ell=|m|=n-1$, denoted $\kcirc{n}$, have  long  lifetimes, in the tens of milliseconds range for $n\approx 50$, which can be even further increased by spontaneous-emission inhibition~\cite{Hulet1985,Nguyen2018,Wu2023,Holzl2024}. Their large electric-dipole matrix elements result in a strong coupling to electromagnetic fields, instrumental for microwave cavity QED~\cite{Haroche2013} and for the metrology of static electric fields~\cite{Facon2016}.

Rydberg-based quantum computation and simulation rely on the strong dipole-dipole interaction between Rydberg atoms~\cite{Saffman2016, Browaeys2020, Morgado2021}. Existing platforms use relatively short-lived and un-trapped laser-accessible low-$\ell$ Rydberg levels. Employing laser-trapped CRAs with a hundred times larger lifetime would considerably improve the performances of Rydberg-based quantum devices~\cite{Nguyen2018,Cohen2021}. Individual Rubidium CRAs have been recently laser-trapped in the ponderomotive potential~\cite{Dutta2000} created by bottle-beam (BoB) optical tweezers~\cite{Ravon2023} with an intensity minimum surrounded by a light shell~\cite{Isenhower2009, Barredo2020}. Strontium CRAs have also been prepared~\cite{Teixeira2020} and laser-trapped in Gaussian optical tweezers addressing an ionic core transition~\cite{Holzl2024}. But dipole-dipole interactions between CRAs have not been observed so far.

In this work, we directly measure these interactions with two Rubidium atoms in the circular Rydberg states $\kcirc{51}$ and $\kcirc{52}$, individually trapped in BoBs at an adjustable distance $d$. 
The first-order dipole-dipole interaction Hamiltonian between the  $\kcirc{n}$ and $\kcirc{(n+1)}$ states reads
\begin{align}
\hat{V}_\mathrm{dd} = V_\mathrm{dd}^{(n)}(\theta, d) \,\hat{\sigma}_X, \quad V_\mathrm{dd}^{(n)}(\theta, d) = C^{(n)}_3 \frac{3\cos^2\theta-1}{d^3}, \label{eq:Vdd}
\end{align}
where $\hat{\sigma}_X$ is the Pauli matrix in the $\{\ket{n\mathrm{C}, (n+1)\mathrm{C}}, \ket{(n+1)\mathrm{C}, n\mathrm{C}}\}$ two-atom basis and $\theta$ is the angle between the interatomic and quantization axes. We use microwave spectroscopy to measure $V_\mathrm{dd}^{(51)}$ as a function of $d$ and of $\theta$. The latter is controlled by the direction of an electric field, which we can tune dynamically.

Moreover, we demonstrate that the interaction between circular states acts as a sensitive probe of the interatomic distance. We use this probe to observe the interaction between the strong static dipoles of low-$\ell$ states involved in the circularization process. The resulting mechanical force induces a sub-$\si{\micro\meter}$-amplitude relative motion between the trapped CRAs revealed by microwave spectroscopy. The observed interaction-induced motion is a clear signature of the coupling between internal and motional degrees of freedom in Rydberg-atom systems~\cite{Mehaignerie2023a, Magoni2023a, Bharti2023}.

\begin{figure*}[t]
 \centering
 \includegraphics[width=.99\linewidth]{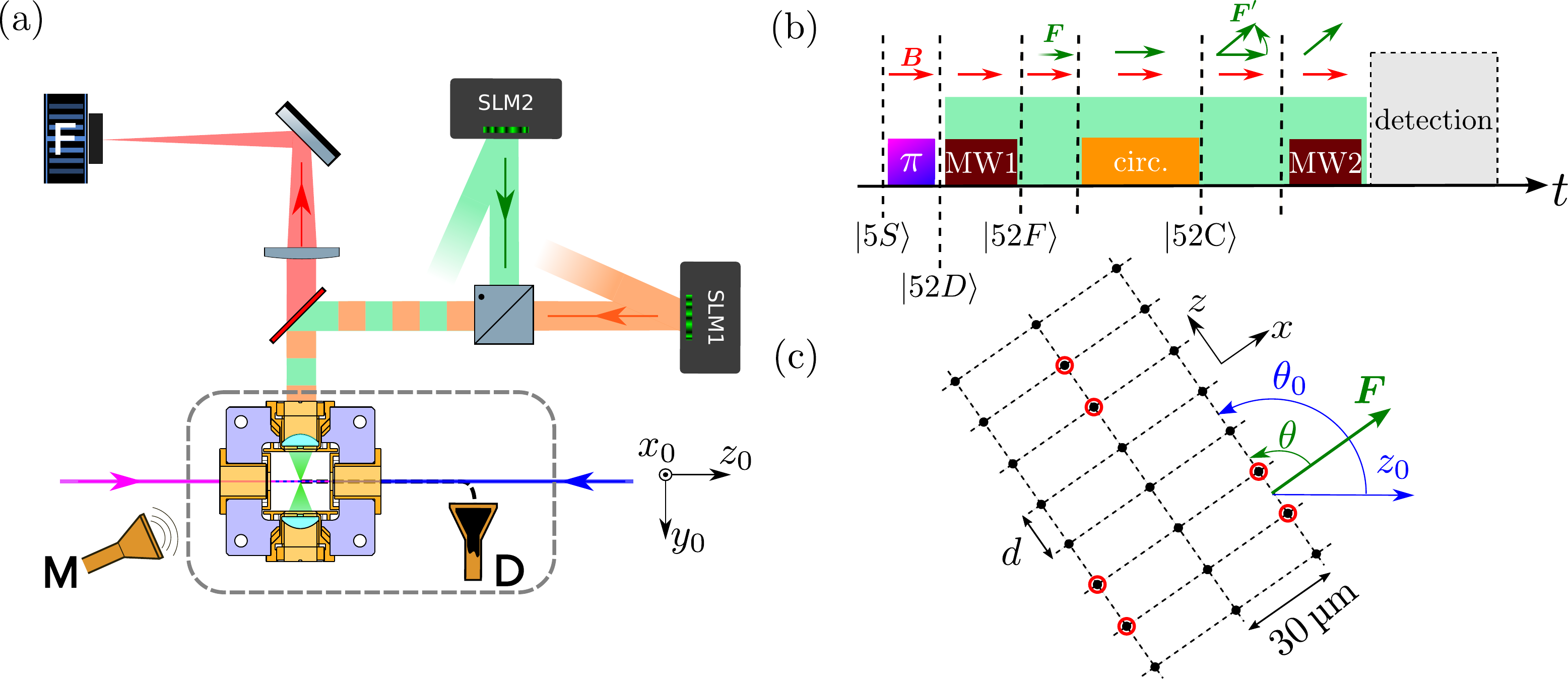}
 \caption{(a) Sketch of the experimental setup: A sapphire cube (violet) holds electric-field control electrodes (gold) and two short-focal-length lenses (cyan). One of them focuses two beams at $\SI{821}{\nano\meter}$ (green and orange). Their phase profiles are tailored by two SLMs. The same lens collects $780$-$\si{\nano\meter}$ light (red) emitted by the trapped atoms. The Rydberg excitation lasers (magenta and blue) enter the cube along the $z_0$ axis. A microwave horn (M) shines a MW field into the UHV chamber (gray dashed square). A channeltron (D) counts Rb ions guided towards it (black dashed line) by electrostatic lenses (not shown). The camera (F) collects the fluorescence photons. (b) Sketch of the experimental sequence (not to scale) following the optical pumping to the $\ket{5S_{1/2}, F=2, m_F=2}$ ground-state. The state of the atom, following the laser excitation $\pi$ pulse (violet rectangle), the MW1 and MW2 pulses (brown) (on the $\ket{52D} \to \ket{52F}$ and $\kcirc{52}\to\kcirc{51}$ transitions, respectively) and the circularization process (orange), is indicated below. The green rectangle indicates the time during which the BoBs are turned on. The arrows depict the relative orientations of the constant magnetic field $\vv{B}$ (red) and of the time-varying electric field $\vv{F}$ (green). Prior to the MW2 pulse, the electric field may be rotated from $\vv{F}$ to $\vv{F}'$. The sequence ends with field-ionization detection (gray) of the Rydberg atoms. (c) Sketch of the trap arrays: The atoms are loaded into the array of Gaussian optical tweezers indicated by black dots. The sites that are circled in red belong to the ``target'' array. The angle $\theta_0$ between the $z$ and $z_0$ axes is constant during one experimental sequence, while the angle $\theta$ between the electric field ($F$) and the $z$ axis can be dynamically tuned.}
 \label{fig:setup}
\end{figure*}

\begin{figure}[t]
 \centering
 \includegraphics[width=.99\linewidth]{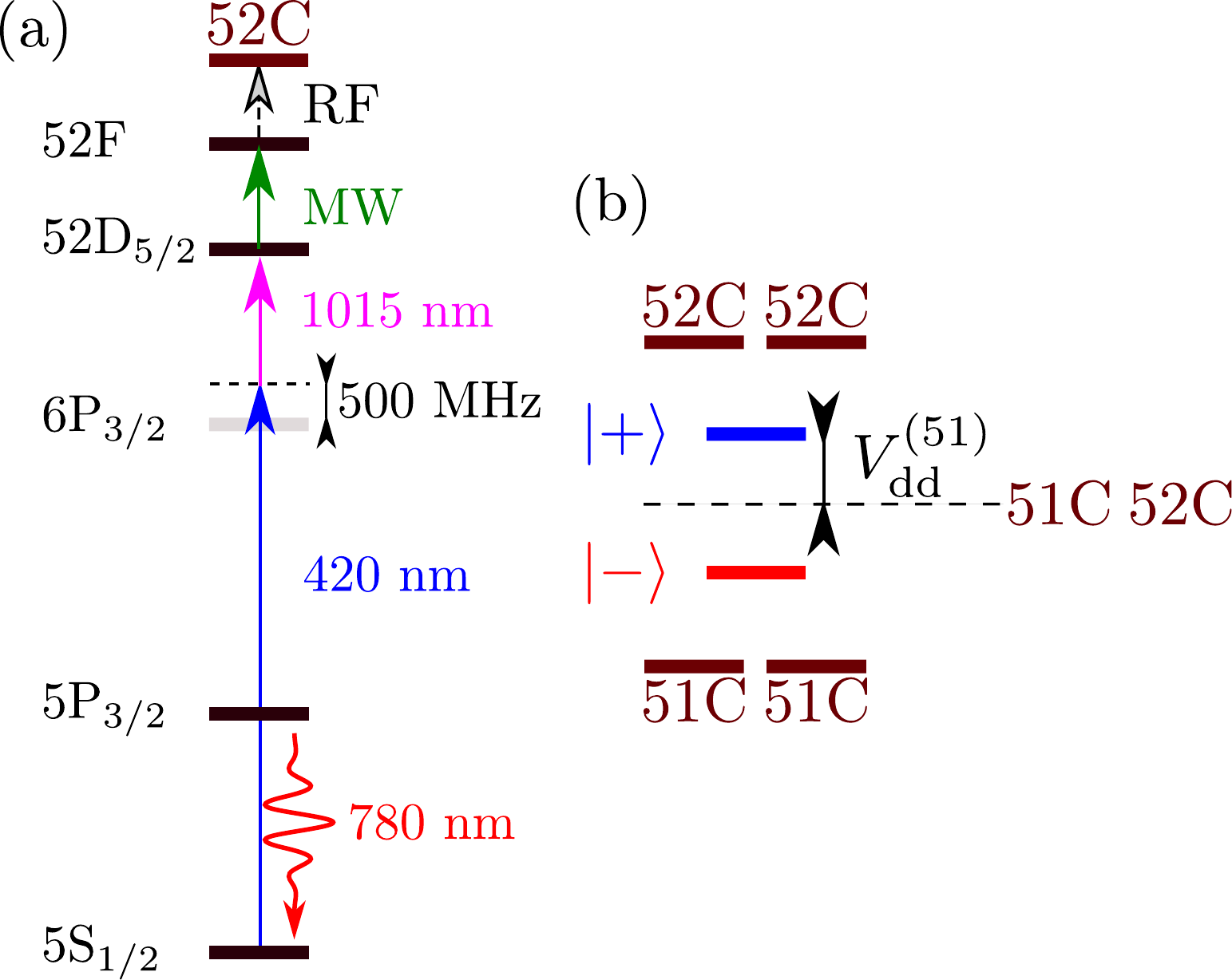}
 \caption{(a) Single-atom level diagram of $^{87}\mathrm{Rb}$ displaying the $\SI{780}{\nano\meter}$ fluorescence line and the Rydberg excitation transitions. (b) Two-atom level diagram in the $\{\kcirc{52},\kcirc{51}\}$ subspace (not to scale). In presence of interaction, the energies of the symmetric $\ket{+}$ and anti-symmetric $\ket{-}$ states ($\ket{\pm}=(\ket{52\mathrm{C}, 51\mathrm{C}} \pm \ket{51\mathrm{C}, 52\mathrm{C}})/\sqrt{2}$) are shifted from the no-interaction levels (dashed line) by $\pm V_\mathrm{dd}^{(51)}$.}
 \label{fig:levels}
\end{figure}

The experimental setup~\cite{Ravon2023}, depicted in Fig.~\ref{fig:setup}, is designed to trap individual ground-state Rubidium-87 atoms in an array of Gaussian optical tweezers~\cite{Schlosser2001} and to promote them into circular Rydberg states trapped in BoBs. The traps are prepared at the focus of a short-focal-length lens by applying dedicated phase masks with two spatial light modulators (SLM1 and SLM2)~\cite{SLM} on a $\SI{821}{\nano\meter}$-wavelength laser. The atoms are excited to circular states by a combination of laser, microwave (MW) and radiofrequency (RF) pulses. The Rydberg excitation lasers enter the cube along the $z_0$ axis (axes definition in Fig.~\ref{fig:setup}).

The initial ground-state optical tweezers form an $8\times3$ array with a $\SI{30}{\micro\meter}$ step along the $x$ direction and a step $d$ along the $z$ direction, set at an angle $\theta_0$ from $z_0$. We load this array with a $\gtrsim\SI{60}{\percent}$ efficiency per site. We record a first fluorescence image to detect the trapped atoms and reorganize them with a moving optical tweezer~\cite{Barredo2016, Endres2016} to prepare the ``target'' array (red circles on Fig.~\ref{fig:setup}) composed of three pairs of atoms. A subsequent image is used to post-select the arrays that have been prepared without missing or superfluous atoms. In each pair, the atoms are separated by the distance $d$, with an interatomic axis along the $z$ axis. The spurious interactions between separate pairs are always negligible (Sec.~\ref{subsec:spurious}). We eventually adiabatically lower the intensity of the optical tweezers to cool down the atoms to $\approx \SI{7}{\micro\kelvin}$.

To excite the atoms into the circular Rydberg state $\kcirc{52}$~\cite{Ravon2023}, we first optically pump them into the $\ket{5S_{1/2}, F=2, m_F=2}$ ground state in a magnetic field $B = \SI{13.9}{\gauss}$ (Sec.~\ref{subsec:MW}) along the $z_0$ axis (Fig.~\ref{fig:setup}), which defines the quantization axis. There is no electric field at this stage [sketch of the experimental sequence in Fig.~\ref{fig:setup}(b)]. We turn off the Gaussian traps and immediately excite the atoms to the $\ket{52D_{5/2}, m_J={5/2}}$ Rydberg level (level diagram in Fig.~\ref{fig:levels}) with a two-photon $\SI{0.5}{\micro\second}$-long  $\pi$-pulse laser excitation. It uses two $\sigma^+$-polarized lasers at $\SI{420}{\nano\meter}$ and at $\SI{1015}{\nano\meter}$. 
We then turn on the array of BoBs, prepared by SLM2, which has the same geometry as the target array and is carefully aligned with it.

A $\SI{0.7}{\micro\second}$-long MW $\pi$ pulse [MW1 in Fig.~\ref{fig:setup}(b)] transfers the atoms into the $\ket{52F, m_F=2} $ state. We ramp up within $\SI{0.9}{\micro\second}$ the electric field $F$ along the $z_0$ axis to $F_0=\SI{1.96}{\volt\per\centi\meter}$. This field ramp adiabatically brings the atoms into a $m=2$ Stark state with a static electric dipole along $z_0$ equal to $3 n (n-5)/2 = 3666$ in atomic units, with $n=52$~\cite{Gallagher1994}. 

To avoid the interaction between these huge dipoles within a pair during the circular-state preparation, we set $\theta_0$ to the ``magic'' value $\theta_m=\arccos (1/\sqrt{3})\approx \SI{54.7}{\degree}$: At this angle, the interaction between static dipoles vanishes, similarly to the interaction between circular states [Eq.~\eqref{eq:Vdd}]. We then increase within $\SI{2}{\micro\second}$ the electric field to $\SI{2.13}{\volt\per\centi\meter}$ while simultaneously shining a $\sigma^+$-polarized RF field at  $\SI{225}{\mega\hertz}$. The atoms are adiabatically transferred into $\kcirc{52}$. The complete circular states preparation lasts $\SI{8}{\micro\second}$ and has a $\gtrsim\SI{70}{\percent}$ efficiency. It is in part limited by the inhomogeneity of the excitation lasers over the whole array and by the room-temperature environment of our setup, which reduces, through blackbody-radiation-induced transfers, the lifetime of $\kcirc{52}$ to $\SI{130}{\micro\second}$~\cite{Ravon2023}. 
We finally switch on the dipole-dipole interactions inside the pairs by adiabatically rotating in the $(x,z)$ plane the electric field, set to $F=\SI{2.05}{\volt\per\centi\meter}$. Since the Stark effect is always an order of magnitude larger than the Zeeman effect, the electric field defines the quantization axis (Sec.~\ref{subsec:MW}). We use its rotation to set $\theta$ to an arbitrary value.

 \begin{figure}[t]
 \centering
 \includegraphics[width=.99\linewidth]{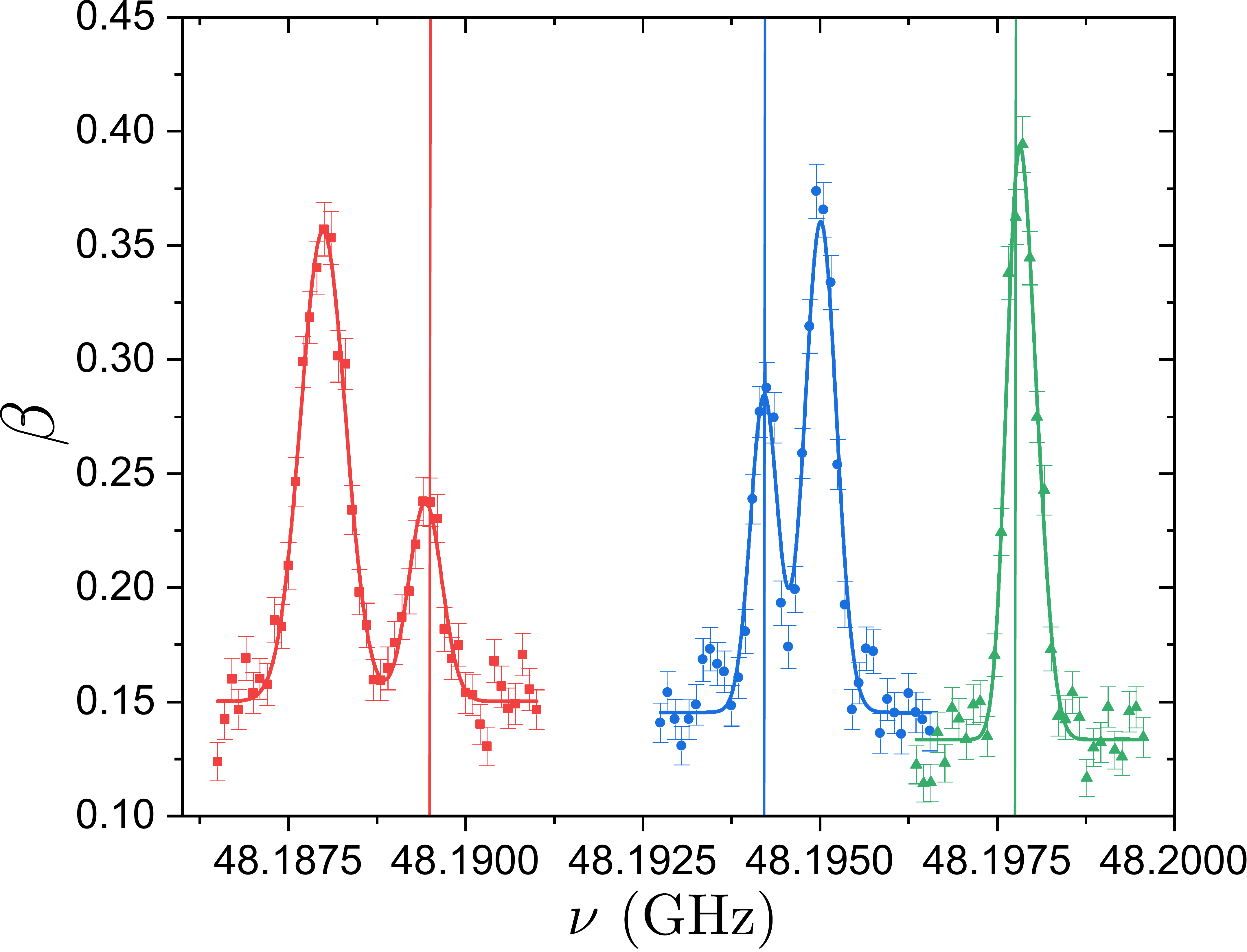}
 \caption{Microwave spectra of the $\kcirc{52}\to\kcirc{51}$ transition. The transfer rate $\beta$ from $\kcirc{52}$ to $\kcirc{51}$ for an atom in a pair is plotted as a function of the frequency $\nu$ of the MW pulse for three different values of $\theta$: $\SI{0}{\degree}$ (red squares), $\SI{90}{\degree}$ (blue circles) and $\SI{60}{\degree}$ (green triangles). The vertical lines indicate the corresponding values of $\nu_0(\theta)$. Every data point results from an average over $400$ realizations of the experiment. Error bars correspond to a 1-$\sigma$ statistical standard-error deviation. The solid lines are fits to the data by a sum of two Gaussian peaks.}
 \label{fig:spectra}
\end{figure}

We measure the interaction between two atoms in $\kcirc{52}$ and $\kcirc{51}$ by microwave spectroscopy. We shine a MW field near resonance with the $\kcirc{52}\to\kcirc{51}$ transition [MW2 in Fig.~\ref{fig:setup}(b)]. For an atomic pair, the MW field couples the initial state $\ket{52\mathrm{C}, 52\mathrm{C}}$ to the symmetric superposition $\ket{+} = (\ket{52\mathrm{C}, 51\mathrm{C}} + \ket{51\mathrm{C}, 52\mathrm{C}})/\sqrt{2}$~\cite{Note1}. The transition to this superposition is shifted from the transition frequency for an isolated atom, $\nu_0(\theta)$, by $\delta\nu(\theta, d) = -V^{(51)}_\mathrm{dd}(\theta, d)/h$, where $h$ is Planck's constant. The frequency $\nu_0(\theta)$ is given by an analytic formula as a function of the angle $(\theta-\theta_0)$ between the magnetic field and the electric field (Sec.~\ref{subsec:MW}). Here, we neglect the second-order van der Waals interactions between atoms in the same circular Rydberg state, which range from $h\times1$ to $\SI{20}{\kilo\hertz}$ in our experiments (Sec.~\ref{sec:theory}).

Setting the distance between the atoms to $d=d_0=\SI{13}{\micro\meter}$, we record the microwave spectrum of the $\kcirc{52}\to\kcirc{51}$ transition by scanning the frequency $\nu$ of a $\approx\SI{2}{\micro\second}$ long MW pulse (Sec.~\ref{subsec:MW}). We measure the population $\Pi_n$ in the $\kcirc{n}$ circular Rydberg states, for $n=52$ and $51$, by field ionization~\cite{Ravon2023}. We plot in Fig.~\ref{fig:spectra} the transfer rate $\beta=\Pi_{51}/(\Pi_{51}+\Pi_{52})$ as a function of $\nu$ for three different values of $\theta$. When $\theta=\SI{0}{\degree}$ or $\SI{90}{\degree}$, two peaks are visible, a clear signature of the inter-atomic interaction. One of them corresponds to the $\ket{52\mathrm{C}, 52\mathrm{C}}\to\ket{+}$ transition. 
The other peak is centered on the frequency $\nu_0(\theta)$, indicated by vertical lines, and corresponds either to the two-photon transition $\ket{52\mathrm{C}, 52\mathrm{C}} \to \ket{51\mathrm{C}, 51\mathrm{C}}$, which is unaffected by the first-order dipole-dipole interactions, or to the situation where one of the two atoms of the pair is missing or has not been properly prepared in $\kcirc{52}$. This happens in particular because of the decay of the circular states during the $\SI{7}{\micro\second}$ delay that separate the state preparation and the MW pulse. We introduce this delay for the electric field to reach a steady value during the MW pulse.

 \begin{figure}[t]
 \centering
 \includegraphics[width=.99\linewidth]{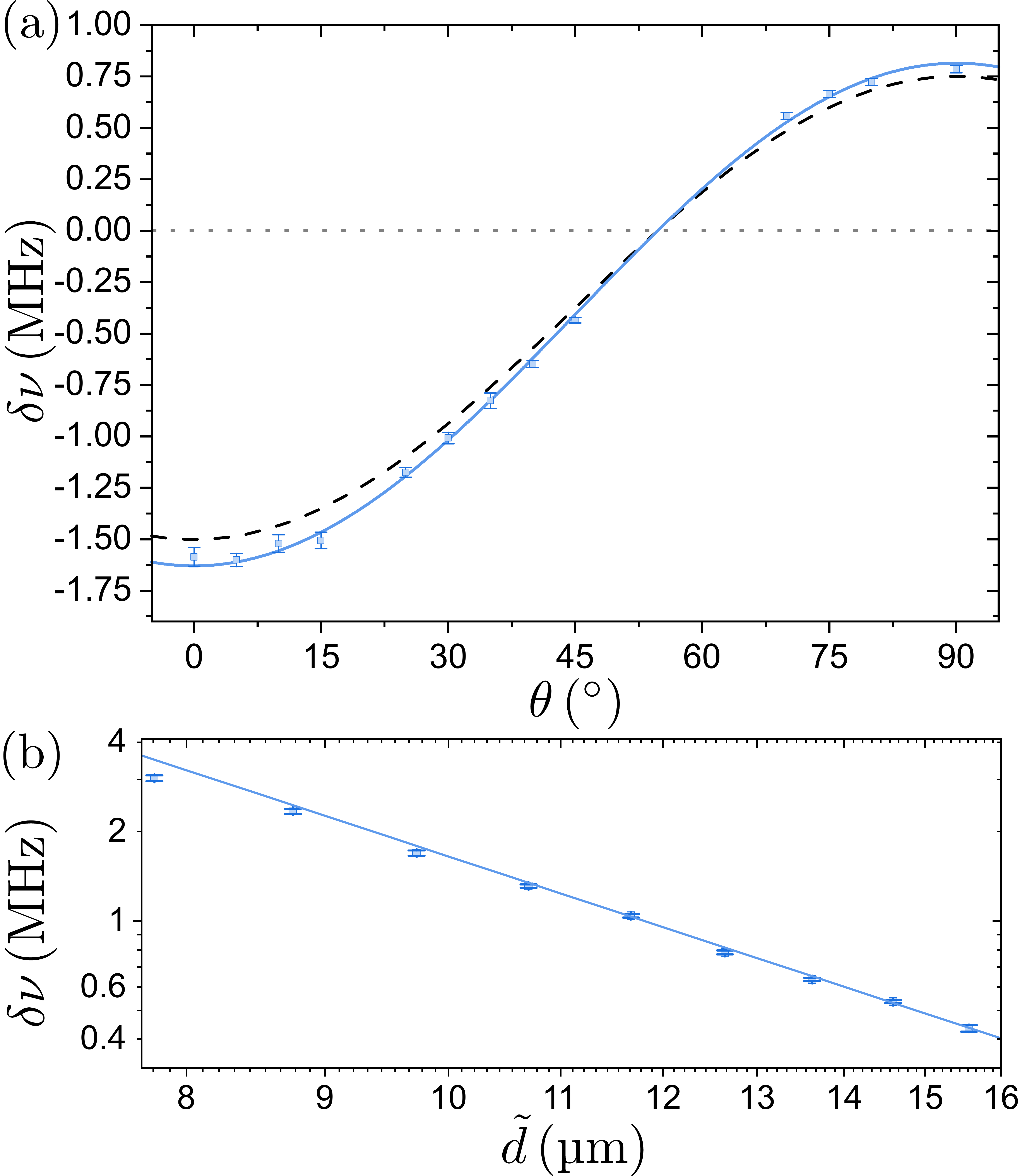}
 \caption{Characterization of the dipole-dipole interaction. (a) The MW line splitting $\delta\nu(\theta, d)$ is plotted as a function of $\theta$ for $d=\SI{13}{\micro\meter}$. The black dashed lines is the prediction of the hydrogenic model (see text) with $d=d_0=\SI{13}{\micro\meter}$. The blue solid line is a fit to the data of the hydrogenic model with $d=\SI{12.65\pm0.04}{\micro\meter}$. The dotted horizontal line is the $\delta\nu=0$ line. (b) The MW line splitting $\delta\nu(\theta, d)$ is plotted as a function of $\tilde{d} = (1-\alpha)d$ for $\theta=\SI{90}{\degree}$. The blue solid line is the prediction of the hydrogenic model. In (a) and (b), every data point results from an average over $160$ realizations of the experiment. Error bars correspond to a 1-$\sigma$ standard error deviation.}
 \label{fig:interactions}
\end{figure}

The collective excitation of the two atoms is confirmed by a measured $\sqrt{2}$ enhancement~\cite{Gaetan2009} of the Rabi frequency on the transition to $\ket{+}$ with respect to the single-atom Rabi frequency on the $\ket{52}\to\ket{51}$ transition (Sec.~\ref{subsec:collective}). Moreover, we find $\delta\nu(\SI{0}{\degree}, d_0)$ and $\delta\nu(\SI{90}{\degree}, d_0)$ to be of opposite signs, as expected from the angular dependency in \eqref{eq:Vdd}. The spectrum recorded with $\theta = \SI{60}{\degree} \approx \theta_m$ exhibits only one peak. It illustrates the strong suppression of the dipole-dipole interaction close to the magic angle.

To obtain a precise value of $\delta\nu(\theta, d_0)$, we fit the spectra to a sum of two Gaussian peaks, as shown in Fig.~\ref{fig:spectra}, and calculate $\delta\nu$ as the difference between the fitted central frequencies. We repeat the procedure for different angles and plot $\delta\nu(\theta, d_0)$ as a function of $\theta$ in Fig.~\ref{fig:interactions}.a. The individual spectra and their fits are given in Sec.~\ref{subsec:MW}. The dashed line corresponds to the analytic prediction of a hydrogenic model $C_{3,\mathrm{H}}^{(n)}=\mathrm{Ry}\,a_0^3\, n^4/2=h\times\SI{1.649}{\giga\hertz\cdot\micro\meter^3}$ with $d=\SI{13}{\micro\meter}$, $n=51$, and where $\mathrm{Ry}$ is the Rydberg unit of energy and $a_0$ the Bohr radius (Sec.~\ref{sec:analytical}). This simple model fully agrees with a more complex numerical estimate that includes the electric-field-, magnetic-field- or interaction-induced level mixing (Sec.~\ref{sec:numerical}).

Theoretical predictions systematically deviate from the measurement results. This can be explained by a mere overestimation of the distance between the atoms. By fitting the hydrogenic model to the data, letting $d$ as the only free parameter, we obtain an excellent agreement for $d = \SI{12.65\pm0.04}{\micro\meter}$. This overestimation of the pair spacing can be traced to an overestimation of the focal length of the tightly focusing lens by only $\alpha = \SI{2.7\pm0.3}{\percent}$, compatible with the lens specifications. In the following, we denote $\tilde{d} = (1-\alpha)\,d$ the rescaled distance between the atoms.

We plot in Fig.~\ref{fig:interactions}(b) in a log-log scale the value of $\delta\nu(\theta=\SI{90}{\degree}, \tilde{d})$. The excellent agreement between experimental data and the theoretical predictions clearly reveals the expected $1/\tilde{d}^3$ dependency of $V_\mathrm{dd}^{(51)}$. The slight deviation for the smallest distance $\tilde{d}=\SI{7.8}{\micro\meter}$ may indicate, besides mere experimental uncertainties, a residual motion of the atoms in their traps. 

We now use the pair interaction as a sensitive probe of the interatomic distance, to measure the motion of the atoms within their BoB traps. Here, we intentionally induce this motion using the strong interaction between the static dipoles of the low-$\ell$ Stark states involved in the circularization process, with $\tilde{d}=\SI{12.65}{\micro\meter}$. In the former experiments, these interactions were canceled by setting $\theta$ to $\theta_m$.

\begin{figure}[t]
 \centering
 \includegraphics[width=.99\linewidth]{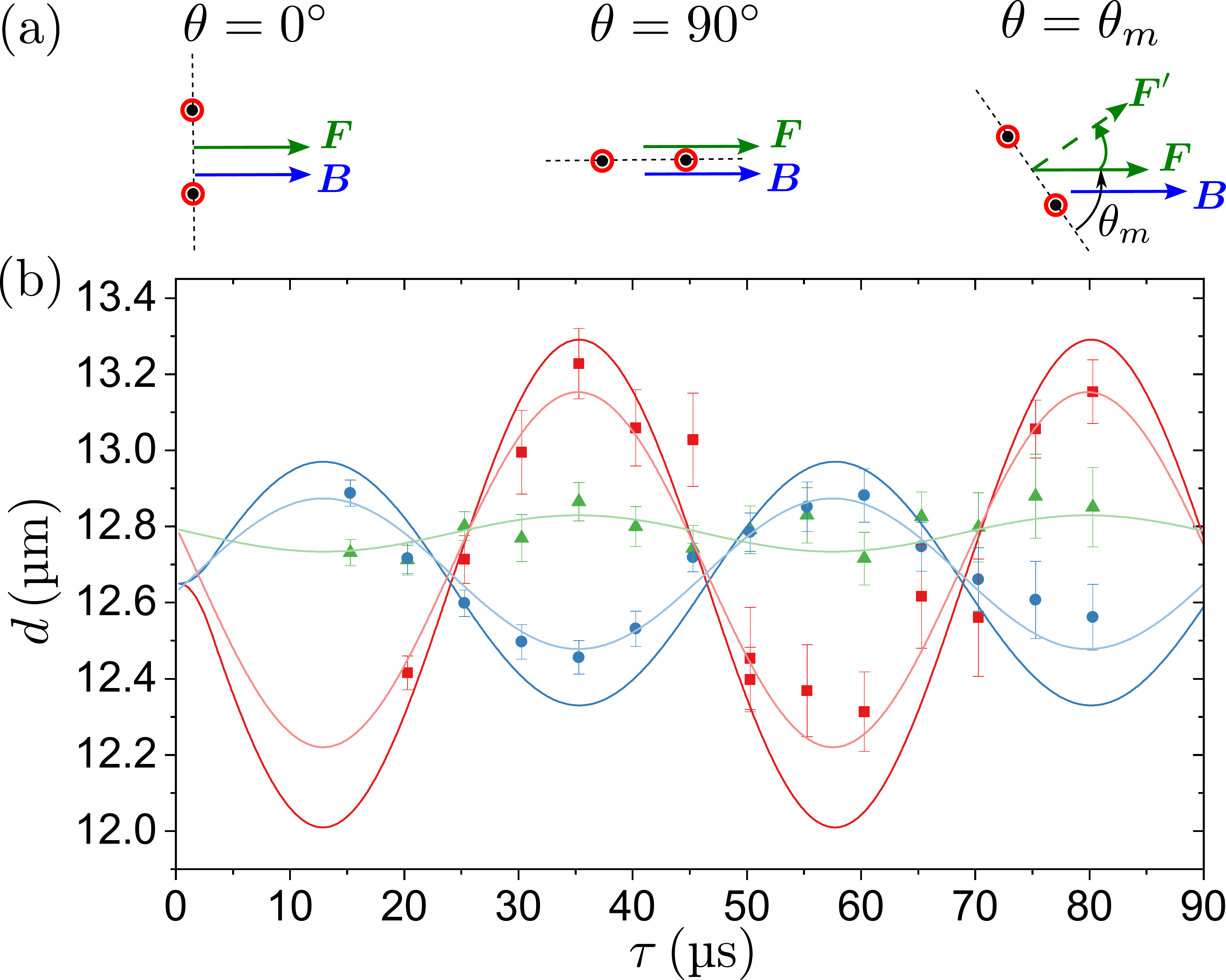}
 \caption{Measurement of the spin-interaction-induced oscillations. (a) Relative orientation of the pair of atoms and the electric and magnetic fields for three values of $\theta$. When $\theta=\theta_m$ during the circularization process, we rotate the electric field from $\vv{F}$ to $\vv{F}'$ after the preparation of $\kcirc{52}$ to set $\theta$ to $\SI{90}{\degree}$. (b) Distance between the circular Rydberg atoms as a function of the time, $\tau$, between the onset of $F_0$ and the MW spectroscopy pulse used to measure $\delta\nu$. The angle $\theta$ is $\SI{0}{\degree}$ (red squares), $\SI{90}{\degree}$ (blue circles) or $\theta_m$ (green triangles). The light lines are sinusoidal fits to the data with amplitudes $\SI{-0.48\pm0.02}{\micro\meter}$, $\SI{0.21\pm0.02}{\micro\meter}$ and $\SI{-0.03\pm0.01}{\micro\meter}$, respectively, and a frequency of $\SI{22.8\pm0.5}{\kilo\hertz}$. The dark lines are the result of an ab-initio calculation (see text). Every data point results from an average over $200$ realizations of the experiment. Error bars correspond to a 1-$\sigma$ standard error deviation.}
 \label{fig:oscillations}
\end{figure}

During the circularization process, the electric and magnetic fields are parallel to the $z_0$ axis. In this situation, the interaction angle $\theta$ is equal to the orientation angle $\theta_0$ of the atomic array. In order to vary the interaction strength between the low-$\ell$ states, we perform three separate experiments with different values of $\theta_0$, setting it to $\SI{0}{\degree}$, $\SI{90}{\degree}$ or $\theta_m$. The three configurations are represented in Fig.~\ref{fig:oscillations}(a). For $\theta=\SI{0}{\degree}$ and $\SI{90}{\degree}$, the atoms are set in motion by the static dipole interaction. It is either repulsive ($\theta>\theta_m$) or attractive ($\theta<\theta_m$). The static dipole vanishes as soon as the circular state has been reached. In the $\theta=\theta_m$ case, there is no interaction between the Rydberg levels. We need, however, to restore the dipole-dipole interactions to measure $\delta\nu$. Therefore, we eventually rotate the electric field, in this latter case only, to set $\theta$ to $\SI{90}{\degree}$ after the circularization process.

From a subsequent measurement of $\delta\nu$, we infer the interatomic distance $d$ and plot it in Fig.~\ref{fig:oscillations}(b) as a function of the time $\tau$ between the setting time of $F_0$ and the MW spectroscopy pulse. The attractive or repulsive character of the force is apparent for $\theta=\SI{0}{\degree}$ and $\SI{90}{\degree}$, respectively. The three datasets are fitted to three sines (light lines) with a common oscillation frequency found to be $\SI{22.8\pm0.5}{\kilo\hertz}$, in fair agreement with the $\SI{24.5}{\kilo\hertz}$ estimated oscillation frequency for the atoms in the BoB traps (Sec.~\ref{subsec:traps}). The fitted amplitudes for $\theta=\SI{0}{\degree}$ and $\theta=\SI{90}{\degree}$ are, with a $\SI{20}{\nano\meter}$ precision, in the few $\SI{100}{\nano\meter}$ range, on the order of the $\lesssim \SI{0.15}{\micro\meter}$-thermal-motion amplitude of an atom in its trap. This confirms the remarkable sensitivity of the dipole-dipole interactions between circular states to the interatomic distance. The amplitudes are found to be in the ratio $-2.3(2)$, close to the expected $-2$ factor. We in addition perform an ab-initio calculation (dark lines), that only takes into account the interaction between the Rydberg Stark states during the adiabatic transfer to $\kcirc{52}$ (Sec.~\ref{sec:simu}). The fair agreement between our data and this numerical simulation confirms the origin of these oscillations. When $\theta=\theta_m$, the motion amplitude significantly reduces to $\SI{-0.03\pm0.01}{\micro\meter}$, a signature of the interaction cancellation. The non-zero value may indicate a mis-alignment with respect to the ideal $\theta=\theta_m$ value by  $\SI{3}{\degree}$ only.  

The observation and characterization of the dipole-dipole interaction between two laser-trapped circular Rydberg atoms is a crucial step towards the realization of quantum simulation~\cite{Nguyen2018} and computation~\cite{Cohen2021} with these atoms. The tools presented here could be readily used in any platform that employs interacting circular Rydberg levels, with alkali or other atoms, such as Ytterbium~\cite{Cohen2021} or Strontium~\cite{Muni2022, Holzl2024}. Analog and digital Rydberg-based quantum computation strongly benefits from a dynamic control over the strength of the interaction~\cite{Schauss2012, Lienhard2018, Ebadi2021, Evered2023}, that we demonstrate here by tuning the orientation and strength of the electric field. This control could be used to induce quenches in a quantum simulation of interacting spin systems~\cite{deLeseleuc2018} or to dynamically turn off or on interactions between separate qubits. Here, we use it to probe the interaction between Rydberg atoms with permanent electric dipoles, through the relative oscillations of the atoms it induces.

Interactions between Rydberg levels with static dipoles have so far been widely overlooked while they could be highly beneficial to Rydberg-based quantum simulation. Experiments on interacting spin-1/2 systems, for which the two spin states are encoded into two Rydberg levels, have been restricted to the simulation of the XY Hamiltonian~\cite{Leseleuc2019, Chen2023}, extended to the simulation of the XXZ Hamiltonian at the expense of Floquet engineering ~\cite{Scholl2022}. Indeed, the first-order spin-exchange-like dipole-dipole interaction usually overwhelms the second-order Ising-like van der Waals interaction between atoms in the same state. As for dipolar molecule systems~\cite{Li2023}, interactions between levels with static dipole would on the contrary provide a tunable first-order Ising-like interaction. This would turn the simulation of an XY Hamiltonian into the simulation of the XXZ one, without the need of dynamic Hamiltonian engineering.

\medskip
This publication has received funding by the France 2030 programs of the French National Research Agency (Grant number ANR-22-PETQ-0004, project QuBitAF), under Horizon Europe programme HORIZON-CL4-2022-QUANTUM-02-SGA via the project 101113690 (PASQuanS2.1), by the European Union (ERC Advanced grant n\textdegree\ 786919, project TRENSCRYBE). It has been supported by Région Île-de-France in the framework of DIM SIRTEQ (project CARAQUES) and by the Quantum Information Center Sorbonne as part of the program investissements d'excellence -- IDEX of the Alliance Sorbonne Universit\'e.

\appendix

\section{Supplementary experimental information}
\label{sec:app_data}

\subsection{Trapping lasers}
\label{subsec:traps}
The geometry of the optical tweezers and of the bottle beams is identical to that described in~\cite{Ravon2023}. The power per optical tweezer is set to $\SI{3.2}{\milli\watt}$ during the atom loading and is later reduced to $\SI{0.6}{\milli\watt}$, prior to the Rydberg state excitation, to adiabatically cool down the atoms to $\approx \SI{7}{\micro\kelvin}$. The power per BoB is $\SI{55}{\milli\watt}$. From the results of~\cite{Ravon2023}, we then expect a $\SI{24.5}{\kilo\hertz}$ transverse oscillation frequency for the circular Rydberg atoms in their traps.

\subsection{Individual microwave spectra}
\label{subsec:MW}

\begin{figure}[t]
\centering
\includegraphics[width=0.99\linewidth]{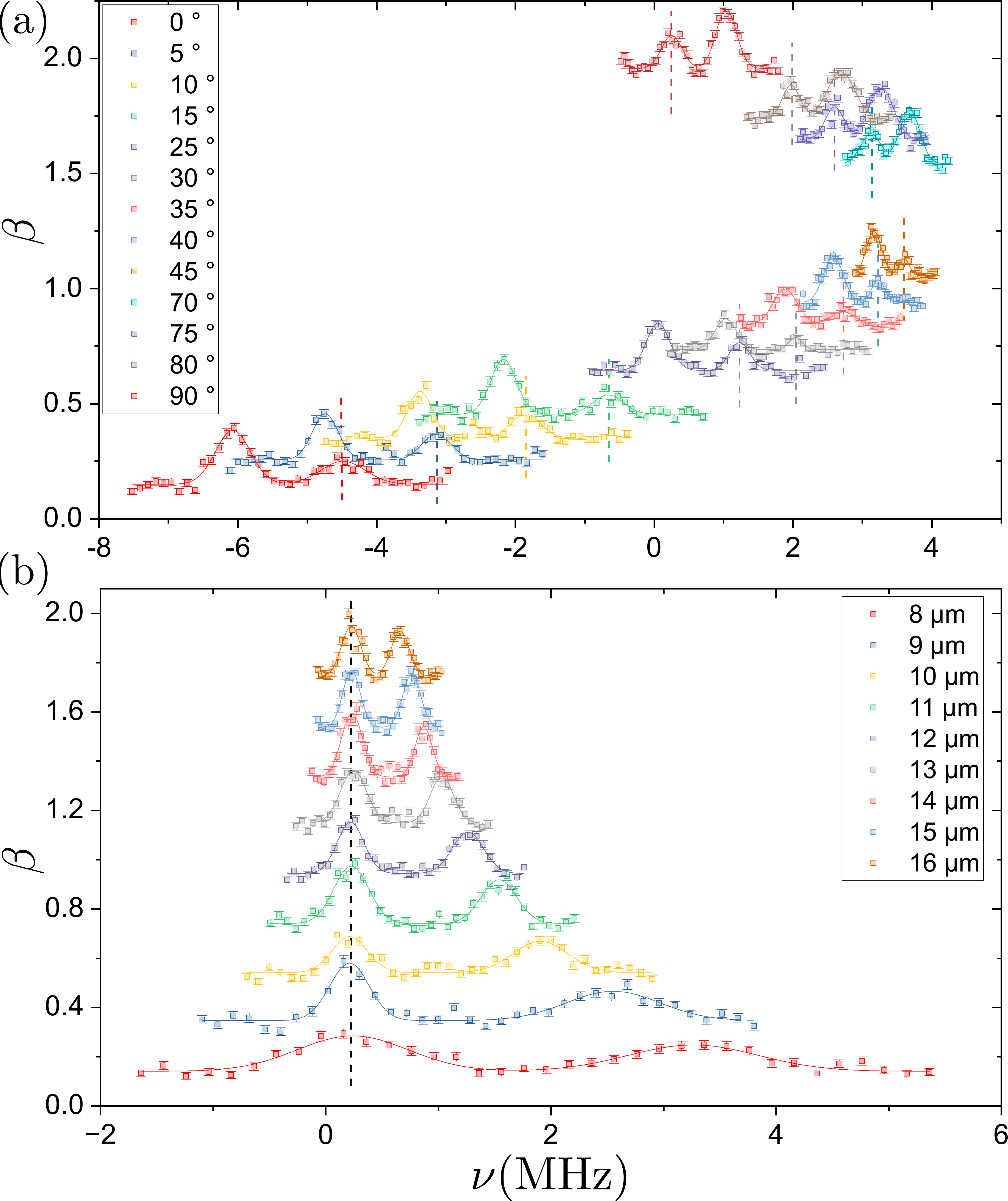}
\caption{Microwave spectra corresponding to the data of Fig.~\ref{fig:interactions}. The transfer rate $\beta$ is plotted as a function of the frequency $\nu$ of of the MW pulse, offset by $\SI{48.194}{\giga\hertz}$, and for different values of $\theta$ (a) and $d$ (b). Their respective values are given in the insets. The individual spectra are displaced along the $y$ axis for visibility by an amount proportional to $\theta$ in (a) and to $d$ in (b). The solid lines are fits to a sum of Gaussian peaks. The dashed lines indicate the peak centered on the isolated-atom resonance frequency $\nu_0(\theta)$.}
\label{fig:individuals}
\end{figure}

In Fig.~\ref{fig:individuals}, we plot the microwave spectra that correspond to the data plotted in Fig.~\ref{fig:interactions}. The microwave spectra are recorded for different values of $\theta$ with $d=\SI{13}{\micro\meter}$ or for different values of $d$ with $\theta=\SI{90}{\degree}$. We use a MW pulse of duration $t_\mathrm{MW}$ given in table~\ref{tab:MWtime}, while the power of the MW field is set to maximize the transfer ratio $\beta$ on the $\ket{52\mathrm{C}, 52\mathrm{C}} \to \ket{+}$ transition.

\begin{figure}[t]
\centering
\includegraphics[width=0.99\linewidth]{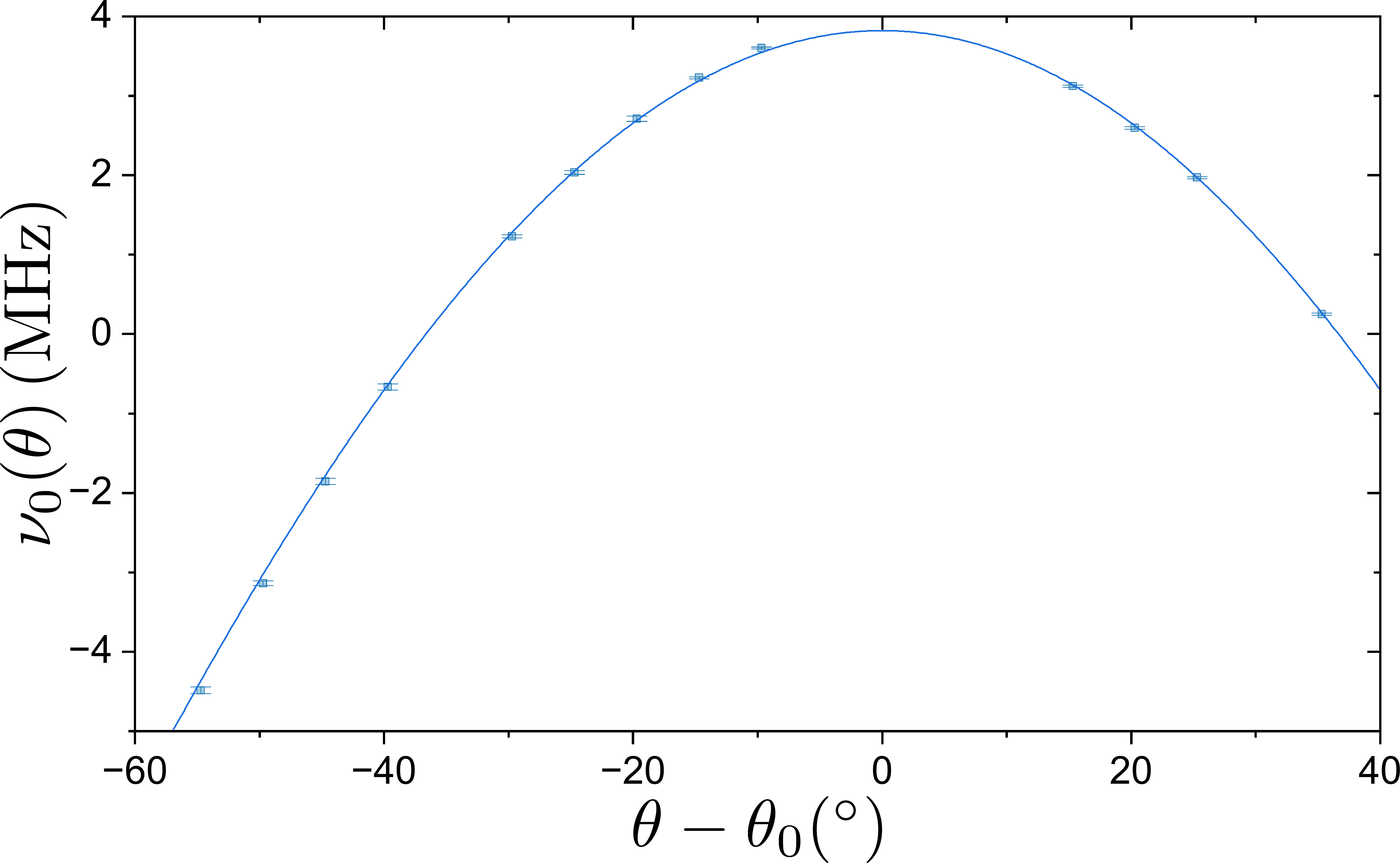}
\caption{Isolated-atom resonance frequency $\nu_0(\theta)$, offset by $\SI{48.194}{\giga\hertz}$, obtained from the microwave spectra of Fig.~\ref{fig:individuals}(a). The solid line is a fit to the theoretical value of~\eqref{eq:nu0} with the magnetic field as the only free parameter. }
\label{fig:zeeman}
\end{figure}

For each spectrum, two peaks are visible. One of them is centered on the isolated-atom resonance frequency $\nu_0(\theta)$ and is marked by vertical dashes lines in Fig.~\ref{fig:individuals}. The measured value of $\nu_0(\theta)$ is plotted in Fig.~\ref{fig:zeeman} as a function of $\theta$. It is in very good agreement with a hydrogenic model that we describe in the following.

The atomic state in the Rydberg manifold of principal quantum number $n$ is described by two angular momenta, $\hat{\vv{J}}_a$ and $\hat{\vv{J}}_b$, with $J_a = J_b = J \equiv (n-1)/2$~\cite{Gallagher1994, Kruckenhauser2022}. The atomic dipole operator $\hat{\vv{D}}$ and the orbital momentum $\hat{\vv{L}}$ read
\begin{align}
	\hat{\vv{D}} = \frac{3}{2\hbar} n\, e a_0\, \left(\hat{\vv{J}}_a - \hat{\vv{J}}_b \right)\ , \quad \hat{\vv{L}} = \frac{1}{2} \left(\hat{\vv{J}}_a + \hat{\vv{J}}_b \right)\ .
\end{align}
In this model, we treat the Zeeman and Stark effects to the first order only. The Stark and Zeeman Hamiltonians, $\hat{v}_Z$ and $\hat{v}_S$, are then cast into the $\hat{v}_a$ and $\hat{v}_b$ Hamiltonians: 
\begin{align}
	\hat{v}_a = (\vv{\Omega_B} - \vv{\Omega_F}) \cdot \hat{\vv{J}}_{a} &,  \quad
	\hat{v}_b = (\vv{\Omega_B} + \vv{\Omega_F}) \cdot \hat{\vv{J}}_{b} \\
	\hat{v}_S + \hat{v}_Z &= \hat{v}_a + \hat{v}_b \ ,
\end{align}
where 
\begin{align}
\vv{\Omega_B} &= \frac{\mu_B}{h}  \vv{B}, \quad 
\vv{\Omega_F} = \frac{3}{2 h} n\, e a_0 \vv{F}.
\end{align}
Here, $\mu_B$ is the Bohr magneton, $e$ the electron charge and $a_0$ the Bohr radius.

\begin{table}[t]
\centering
$\begin{array}{|c|c||c|c|}
\theta (\si{\degree}) & t_\mathrm{MW} (\si{\micro\second}) & d (\si{\micro\meter}) & t_\mathrm{MW} (\si{\micro\second}) \\
\hline
0 & 1.3 & 8 & 0.9 \\
5 & 1.5 & 9 & 1.6 \\
10 & 1.6 & 10 & 2.1 \\
15 & 1.7 & 11 & 2.3 \\
25 & 2.1 & 12 & 2.7 \\
30 & 2.6 & 13 & 3.1 \\
35 & 2.1 & 14 & 3.8 \\
40 & 2.4 & 15 & 4.3 \\
45 & 3.1 & 16 & 4.6 \\
70 & 2.3 &  & \\
75 & 2 &  & \\
80 & 1.8 &  & \\
85 & 1.8 &  & \\
90 & 2.1 &  &
\end{array}$
\caption{Duration, $t_\mathrm{MW}$, of the MW pulse used to record the spectra of Fig.~\ref{fig:individuals}. The left columns correspond to the data of Fig.~\ref{fig:individuals}(a) (fixed $d$, variable $\theta$), the right columns to the data of Fig.~\ref{fig:individuals}(b) (variable $d$, fixed $\theta$).}
\label{tab:MWtime}
\end{table}

The circular Rydberg state is the state $\kcirc{n}=\ket{J_a=J, m_a=J; J_b=J, m_b=J}$, where $\hbar m_a$ ($\hbar m_b$) is the projection of $\hat{\vv{J}}_a$ ($\hat{\vv{J}}_b$) onto the direction of $\vv{\Omega_F} - \vv{\Omega_B}$ ($ \vv{\Omega_F} + \vv{\Omega_B}$). Its eigenenergy related to the Hamiltonian $\hat{v}_Z+\hat{v}_S$ then reads
\begin{align}
\nu(n, \theta) &= \frac{n-1}{2} \left[ \norm{\vv{\Omega_F} + \vv{\Omega_B}} - \norm{\vv{\Omega_F} - \vv{\Omega_B}} \right] \nonumber \\
& = \frac{n-1}{2} \left[\sqrt{\Omega_B^2 + \Omega_F^2 + 2 \Omega_B \Omega_F \cos(\theta)}\right. \nonumber \\
&- \left.\sqrt{\Omega_B^2 + \Omega_F^2 - 2 \Omega_B \Omega_F \cos(\theta)} \right]\ . \label{eq:zeeman}
\end{align} 

From this model, we deduce as a function of the angle $\theta$ the expected frequency of the $\kcirc{52}\to\kcirc{51}$ transition to which we fit the measured frequencies plotted in Fig.\ref{fig:zeeman}:
\begin{align}
\nu_0(\theta) = \overline{\nu}_0(F) + \nu(n=52, \theta) - \nu(n=51, \theta)\ , \label{eq:nu0}
\end{align} 
where the frequency $\overline{\nu}_0(F)$ is the frequency of the $\kcirc{52}\to\kcirc{51}$ transition when $B=0$. For a better precision of the model, we take into account in the estimation of $\overline{\nu}_0(F)$ the second-order Stark shift. The $B=0$ frequency reads~\cite{Gallagher1994}, with $n=52$,
\begin{align}
\overline{\nu}_0(F) &= \frac{\mathrm{Ry}}{h n^2}\left[\frac{2n-1}{(n-1)^2}\right. \nonumber \\
					& \left.- \frac{(h \Omega_F)^2}{\mathrm{Ry}^2} \frac{24 n^5 - 15 n^4 + 10 n^3 - n}{36}\right].
\end{align}
The fit of the data in Fig.~\ref{fig:zeeman} is made by fixing the electric field to $F = \SI{2.05}{\volt\per\centi\meter}$ and letting $B$ as the only free parameter. We find a very good agreement with $B=\SI{13.91\pm0.01}{\gauss}$. With $F=\SI{2.13}{\volt\per\centi\meter}$, this corresponds to $\Omega_F = \SI{1.96E8}{\radian\per\second}$ and $\Omega_B = \SI{1.946E7}{\radian\per\second}$ so that $\Omega_F \gg \Omega_B$. This justifies why we can consider the quantization axis to be aligned with the electric field whatever the value of $\theta$.

\begin{figure}[t]
\centering
\includegraphics[width=0.99\linewidth]{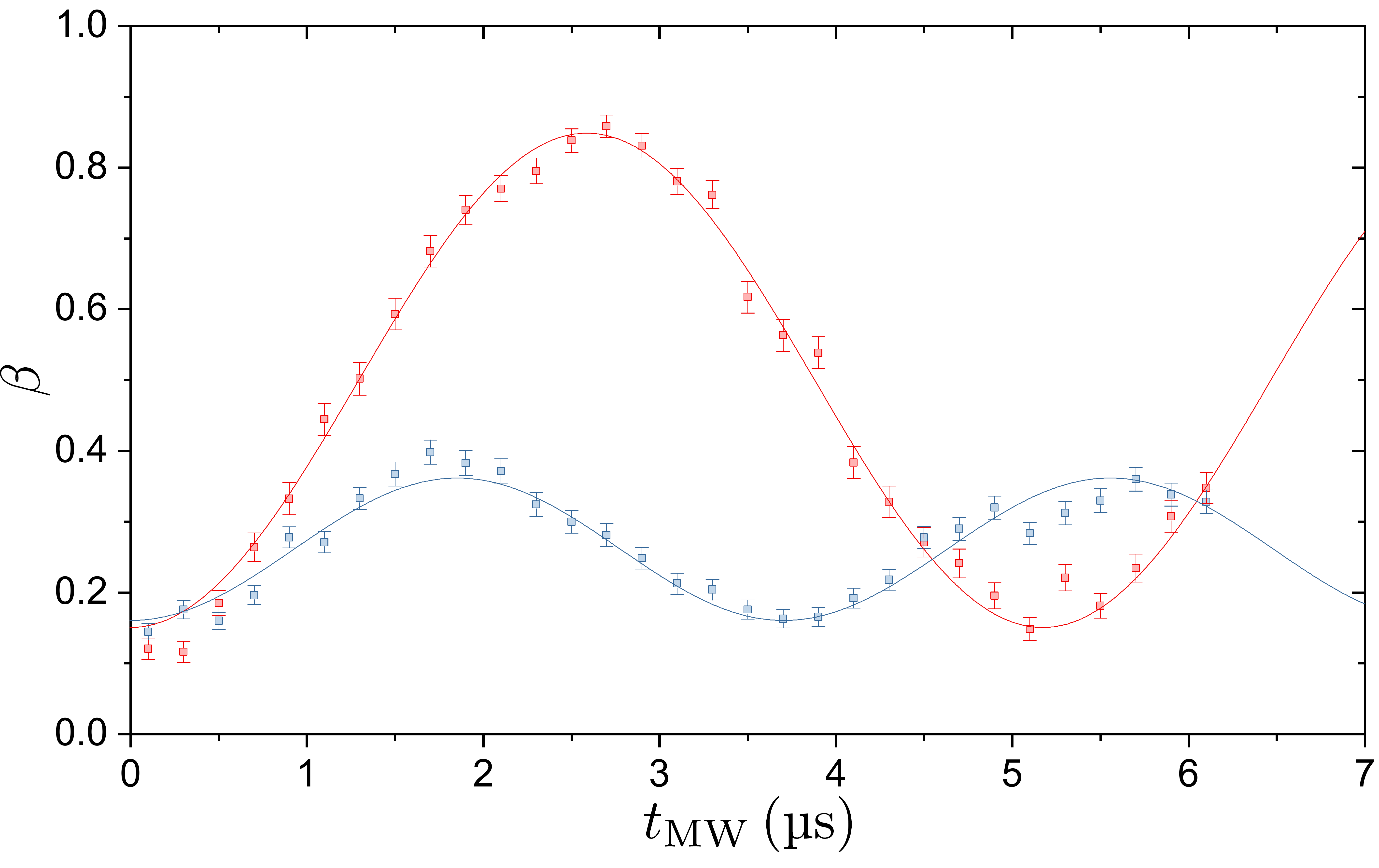}
\caption{Transfer rate $\beta$ measured as a function of the duration of the MW pulse, $t_\mathrm{MW}$, that drives Rabi oscillations recorded on the $\kcirc{52}\to\kcirc{51}$ transition with a single atom (red squares) or on the $\ket{52\mathrm{C}, 52\mathrm{C}}\to\ket{+}$ transition with a pair of atom. The solid lines are fits of the data to a sine law.}
\label{fig:Rabi}
\end{figure}

\subsection{Collective coupling}
\label{subsec:collective}
To confirm the excitation of the symmetric two-atom state $\ket{+}$, we record Rabi oscillations on either the $\ket{52\mathrm{C}, 52\mathrm{C}} \to \ket{+}$ or the $\kcirc{52}\to\kcirc{51}$ transition. The former is recorded with a pair of atoms, in the same conditions as for the measurement of the microwave spectra of Fig.\ref{fig:individuals}(a), with $\theta=\SI{90}{\degree}$, setting the MW-field frequency to $\nu_0(\theta) + \delta\nu(\theta, \tilde{d} = \SI{12.65}{\micro\meter})$. The latter is recorded by preparing only one atom in each pair to the circular state $\kcirc{52}$, setting the MW field frequency to $\nu_0(\theta)$. In Fig.~\ref{fig:Rabi}, we plot $\beta$ as a function of the duration $t_\mathrm{MW}$ of the MW pulse, in red for a single atom and in blue for a pair of atoms excited to $\ket{+}$. A fit to the data of a sine law reveals the Rabi frequencies $\Omega_1 = 2\pi\times\SI{193\pm1}{\kilo\hertz}$ and $\Omega_2 = 2\pi\times\SI{270\pm3}{\kilo\hertz}$, for the single and two-atom cases, respectively. This corresponds to a ratio $\Omega_2/\Omega_1 = \SI{1.40\pm0.02}{}$, in perfect agreement with the expected $\sqrt{2}$ factor, signature of the collective excitation~\cite{Gaetan2009}.

\subsection{Interactions between different pairs}
\label{subsec:spurious}
\begin{figure}[t]
\centering
\includegraphics[width=0.99\linewidth]{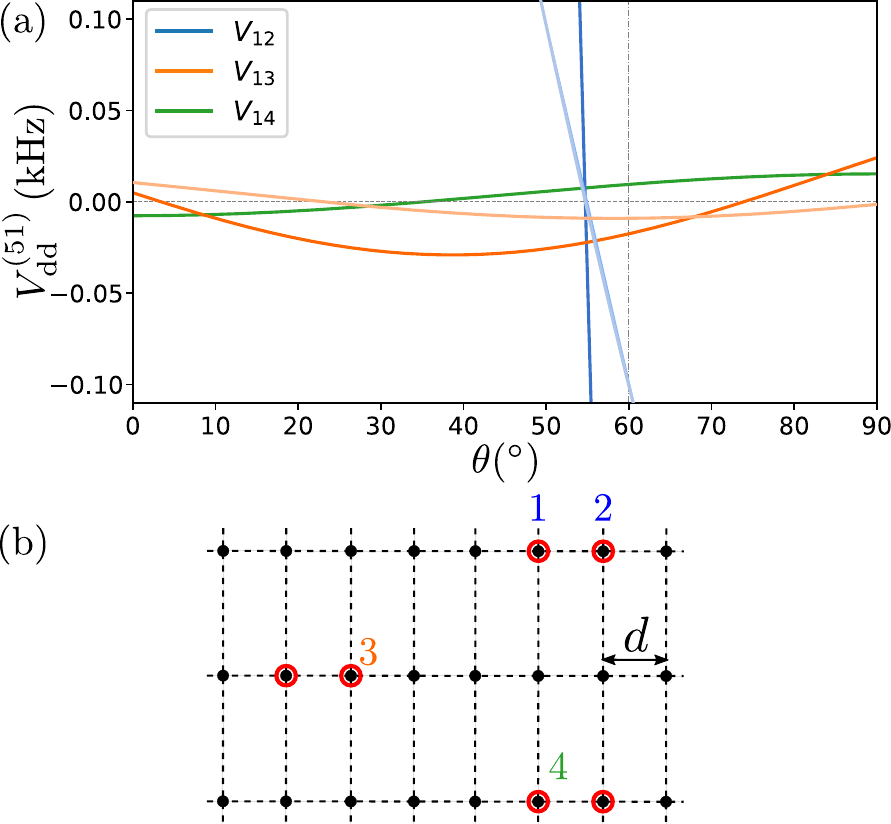}
\caption{(a) Interaction strength $V_\mathrm{dd}^{(51)}(\theta, d)$ between two atoms within a pair (blues lines) compared to interaction strengths between atoms in separate pairs. The green line corresponds to the interaction between atom 1 and atom 4, as labeled in the array plotted in (b). Orange lines correspond to the interaction between atom 1 and atom 3. The dark lines are plotted for $d=\SI{8}{\micro\meter}$ and the light lines for $d=\SI{16}{\micro\meter}$. The vertical dashed line indicates the value $\theta=\SI{60}{\degree}$ corresponding, in particular, to the green spectrum in Fig.~\ref{fig:spectra}.}
\label{fig:Spurious}
\end{figure}

In our setup, we prepare three pairs of interacting circular Rydberg atoms. The relative orientations of the interacting atoms is given by the target array, plotted in Fig.~\ref{fig:setup}(c) and reproduced in Fig.~\ref{fig:Spurious}(b). We focus on the interactions between two atoms in the same pair, e.g., the atoms labeled $1$ and $2$ in Fig.~\ref{fig:Spurious}(b). To estimate the influence of the atoms from the other pairs, we plot in Fig.~\ref{fig:Spurious}(a) the strength of the interaction between atom 1 and atom 2 ($V_{12}$), atom 1 and atom 3 ($V_{13}$), and atom 1 and atom 4 ($V_{14}$) as a function of $\theta$, using the predictions of the hydrogenic model.

We note that the spurious inter-pair interactions $V_{13}$ and $V_{14}$ are in all considered situations weaker than $h\times\SI{30}{\kilo\hertz}$, i.e., negligible with respect to the measured interactions under the conditions considered in the main text. The inter-pair interactions become non-negligible only close to the magic angle $\theta_m$, for which the intra-pair interaction vanishes. At $\theta=\SI{60}{\degree}$, however, $|V_{12}|\gg|V_{13}|, |V_{14}|$ for all considered values of $d$.

\section{Numerical simulation of the interaction-induced relative motion}
\label{sec:simu}

In this section, we describe the semi-classical model used to simulate the relative motion of the trapped circular Rydberg atoms, shown in Fig.~\ref{fig:oscillations}. We consider two atoms, each trapped in a one-dimensional harmonic potential approximating the trapping potential of the BoBs. The trapping frequency is set to the measured value of $\SI{22.8}{\kilo\hertz}$ and the atomic motion is treated classically.

We take into account the interactions between Rydberg atoms only after the electric field has been set to its non-vanishing value $F_0$. We denote $\tau$ the time that originates at the onset of $F_0$. The atoms interact during the whole circularization process. At long times, the atoms are in the same circular Rydberg states and only interact through second-order van der Waals interaction, that we neglect. To calculate the interaction strength during the circularization process, we use the hydrogenic model introduced in Sec~\ref{subsec:MW}.

During the circularization process, the electric and magnetic fields are parallel to a common unit vector $\vv{n}$, which defines the quantization axis. In this situation, $m = m_a + m_b$, and the $\hat{v}_a$ and $\hat{v}_b$ Hamiltonians simplify to 
\begin{align}
	\hat{v}_a = (\Omega_B - \Omega_F)  \hat{\vv{J}}_{a} \cdot \vv{n}\ , \quad \hat{v}_b =  (\Omega_B + \Omega_F)  \hat{\vv{J}}_{b} \cdot \vv{n}\ .
\end{align}
The circular Rydberg state is the state $\kcirc{n}=\ket{J_a=J, m_a=J; J_b=J, m_b=J}$, while the initial state is the $m=0$ state $\ket{J_a=J, m_a=J, J_b=J, m_b=-J}$. In the hydrogenic model, these two states are coupled through the adiabatic circularization process. Note that this slightly differs from our experiment with $^{87}\mathrm{Rb}$ atoms where $\kcirc{n}$ is coupled to a $m=2$ state.

Under the action of the $\sigma^+$-polarized RF field, of angular frequency $\omega_\mathrm{RF}$, the angular momentum $\hat{\vv{J}}_b$ rotates from $m_b = -J$ to $m_b = J$. This process is described, under the rotating wave approximation, by the Hamiltonian
\begin{align}
	\hat{H}_\mathrm{RF} = \Omega_b \left(\hat{J}_{b,+}\, \e^{i \omega_\mathrm{RF} t} + \hat{J}_{b,-}\, \e^{-i \omega_\mathrm{RF} t}\right).
\end{align}
In the frame where $\hat{\vv{J}}_b$ rotates at $\omega_\mathrm{RF}$, considering both $\vv{F}$ and $\vv{B}$ to be aligned along the $z$ axis, the Hamiltonian that governs the evolution of the state of an isolated atom then reads
\begin{align}
	\hat{H} = -\omega_a \hat{J}_{a,z} - \Delta_b \hat{J}_{b,z} + \Omega_b \left(\hat{J}_{b,+} + \hat{J}_{b,-}\right) \ ,
\end{align}
where $\omega_a = \Omega_F - \Omega_B$ and $\Delta_b = \omega_\mathrm{RF} -  (\Omega_B + \Omega_F)$. In our experiment, with $B = \SI{13.9}{\gauss}$, we have $\Delta_b = 0$ when $F = \SI{2.06}{\volt\per\centi\meter}$.

For a pair of atoms, we include the dipole-dipole interaction Hamiltonian, which can be written as a function of the angular momenta $\hat{\vv{J}}_{a}^{(i)}$ and $\hat{\vv{J}}_{b}^{(i)}$ of atom $i, i\in\{1,2\}$. It reads~\cite{Kruckenhauser2022}
\begin{align}
\hat{V}_\mathrm{dd} &= \frac{1}{4\pi\varepsilon_0 d^3} \left(\frac{3}{2} n e a_0\right)^2 \left\{\left(\hat{\vv{J}}_{a}^{(1)}-\hat{\vv{J}}_{b}^{(1)}\right)\cdot\left(\hat{\vv{J}}_{a}^{(2)}-\hat{\vv{J}}_{b}^{(2)}\right) \right. \nonumber \\
&\left. - 3 \left[ \left(\hat{\vv{J}}_{a}^{(1)}-\hat{\vv{J}}_{b}^{(1)}\right) \cdot \vv{u}\right] \left[ \left(\hat{\vv{J}}_{a}^{(2)}-\hat{\vv{J}}_{b}^{(2)}\right) \cdot \vv{u}\right] \right\}\ ,
\end{align}
where $\vv{u}$ is the unit vector that points along the interatomic axis. Using Ehrenfest theorem, we get the differential equations that govern the time evolution of the average values of the components of ${\vv{J}}_{a}^{(1)}$, ${\vv{J}}_{a}^{(2)}$, ${\vv{J}}_{b}^{(1)}$ and ${\vv{J}}_{b}^{(2)}$, under the total Hamiltonian $\hat{H}+\hat{V}_\mathrm{dd}$. The angular momenta of the two atoms are initialized to the same values, $\vv{J}_a = \vv{J}_b = (0, 0, -J)$. The two atoms are then in the same state during the whole circularization process: ${\vv{J}}_{a}^{(1)}(t) = {\vv{J}}_{a}^{(2)}(t)$ and ${\vv{J}}_{b}^{(1)}(t) = {\vv{J}}_{b}^{(2)}(t)$ . A numerical integration of the equations then provides us the time-evolution of the electric dipoles of the two atoms $\vv{D}^{(1)}(t) = \vv{D}^{(2)}(t)$ during the circularization process and their dipole-dipole interaction $V(t)$.

We eventually obtain the relative motion $d(t)$ between the two atoms by integrating the classical equation of motion
\begin{align}
	\frac{M}{2} \dderiv{d}{t} = - \frac{M}{2} \omega_t^2 d(t) - \deriv{V(t)}{d} \ .
\end{align}
The values of $d(t)$, obtained for three different values of $\theta$, are plotted in Fig.~\ref{fig:oscillations}.

\section{Theoretical characterization of the interaction
  between two atoms in circular Rydberg states}
\label{sec:theory}

We consider  ${}^{87}\mathrm{Rb}$ atoms in  (nearly-)circular Rydberg states. These are described using the model of a spinless hydrogen atom, whose quantum states $\ket{n,l,m}$ are labelled by the principal, orbital, and magnetic quantum numbers $n$, $l$, $m$, where $n\gtrsim 50$ and $l$, $m$ are close to $n-1$. The two leading corrections to this model, both neglected, are:
\begin{itemize}
\item the quantum defect \cite[chap.~16]{Gallagher1994}, proportional to $1/n^8$ and of the order of $-h\times 1\,\mathrm{kHz}$,
\item the fine-structure splitting \cite[\S    34]{landau4:BH1982}, proportional to $1/n^5$ and of the order of $h\times0.5\,\mathrm{kHz}$.
\end{itemize}
Hence, the calculated energies are significant up to a few $\mathrm{kHz}$.

The Hamiltonian  $\hat{h}^{(i)} = \hat{h}_0^{(i)} + \hat{v}_S^{(i)} + \hat{v}_Z^{(i)}$ for  atom $i$, $i\in\{1,2\}$, includes the bare-atom terms $\hat{h}_0$, and the Stark and Zeeman couplings $\hat{v}_S$ and $\hat{v}_Z$ to the  external electric field $\vv{F}$ and magnetic field $\vv{B}$, respectively.  The distance  $d\gtrsim \SI{8}{\micro\meter}$ between the two atoms greatly exceeds their size $n^2a_0 = \SI{0.14}{\micro\meter}$, so they interact via the dipole-dipole (dd) interaction:
\begin{equation}
  \label{eq:VDD}
  \hat{V}_\mathrm{dd}(\vv{r}_1,\vv{r}_2)/E_\mathrm{H}=
  a_0\: [
    \vv{r}_1\cdot\vv{r}_2
    -3(\vv{r}_1\cdot\vv{u})(\vv{r}_2\cdot\vv{u})
  ]/d^3
  \ ,
\end{equation}
with $E_\mathrm{H} = 2\mathrm{Ry}$ being  the Hartree energy,
$\vv{r}_i$ the  position of the  electron of atom $i$,  and $\vv{u}$
the unit vector pointing along the internuclear axis.

The frequency difference $h \delta\nu(\theta)$ measured experimentally corresponds to the difference between the frequency of 
the $\ket{52\mathrm{C}, 52\mathrm{C}}\to\ket{+}$ transition and the frequency of either
\begin{itemize}
\item[$(I)$] the single-photon
transition bringing  a single atom from $\kcirc{52}$  to $\kcirc{51}$,
\item[$(II)$] or the two-photon transition  bringing two interacting atoms
from  $\ket{52\mathrm{C},52\mathrm{C}}$ to  $\ket{51\mathrm{C},51\mathrm{C}}$.   
\end{itemize}
The corresponding frequency  differences $\delta \nu_I$ and $\delta\nu_{II}$ read:
\begin{align}
  \label{eq:DeltaEI_DeltaEII}
  h \delta\nu_I &= \Delta E_{I} = \delta E_{52\mathrm{C},52\mathrm{C}} - \delta E_{51\mathrm{C},52\mathrm{C}} \ ,\\
  h \delta\nu_{II} &= \Delta E_{II} =
  \frac{\delta E_{52\mathrm{C},52\mathrm{C}}+\delta E_{51\mathrm{C},51\mathrm{C}}}{2}-\delta E_{51\mathrm{C},52\mathrm{C}}
  \ .
\end{align}
Here, $\delta E_{n_1\mathrm{C},n_2\mathrm{C}}=E_{n_1\mathrm{C},n_2\mathrm{C}}-(\epsilon_{n_1\mathrm{C}}+\epsilon_{n_2\mathrm{C}})$ is the difference between the  two-atom energy $E_{n_1\mathrm{C},n_2\mathrm{C}}$ that includes interactions, and the sum  of the energies  $(\epsilon_{n_1\mathrm{C}}+\epsilon_{n_2\mathrm{C}})$ of two non-interacting atoms in the states $\kcirc{n_1}$ and $\kcirc{n_2}$, dressed by $\hat{v}_Z$ and $\hat{v}_S$. The two-atom energies $E_{n\mathrm{C},n\mathrm{C}}, n\in\{51,52\}$ and $E_{52\mathrm{C},51\mathrm{C}}$ are the energies of the $\ket{n\mathrm{C}, n\mathrm{C}}$ and $\ket{+}$ states, respectively, dressed by $\hat{v}_Z$, $\hat{v}_S$ and $\hat{V}_\mathrm{dd}$.  For all considered values of $\vv{B}$, $\vv{F}$ and $d$, $\Delta E_{I}$ and $\Delta E_{II}$ are nearly equal.

\begin{figure}
  \centering
  \includegraphics[width=.99\linewidth]
  {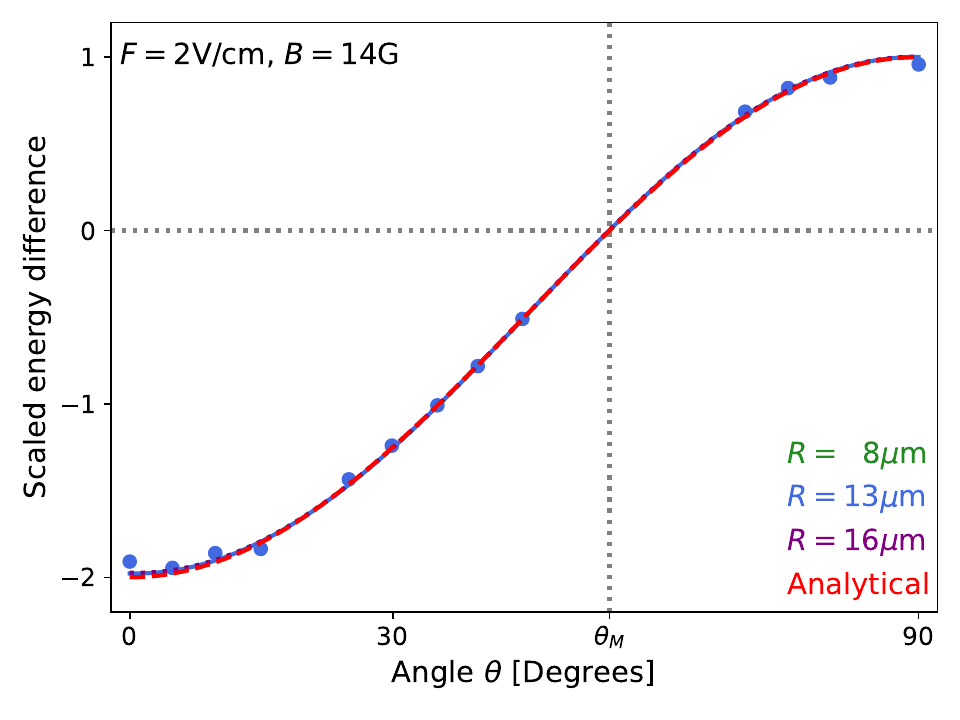}
  \caption{ \label{fig:scaledEnergy}
    Scaled energy difference $\Delta E_I/( C_{3,\mathrm{H}}^{(51)} /d^3)$,
    calculated using the approach of Sec.~\ref{sec:numerical}
    for $d=8$ (green curve), $13$ (blue curve), and
    $\SI{16}{\micro\meter}$ (purple curve). The three curves
    collapse onto the universal curve $(1-3\cos^2\theta)$
    expected from Eq.~\eqref{eq:VDD0} (red curve). All are in excellent agreement with the experimental points from Fig.~\ref{fig:interactions}(a), rescaled by $ C_{3,\mathrm{H}}^{(51)}/\tilde{d}^3$.
    ($\tilde{d}=\SI{12.65}{\micro\meter}$, blue dots).
  }
\end{figure}

We have evaluated $\Delta E_{I,II}$ using two independent approaches: \textit{(i)} an analytical one, described in Sec.~\ref{sec:analytical}, and \textit{(ii)} a numerical  one, presented in  Sec.~\ref{sec:numerical}.  Neither approach involves fits or adjustable parameters. The results are compared to experimental data in Fig.~\ref{fig:scaledEnergy}.

\subsection{\label{sec:analytical}
  Analytical expression for $\Delta E_{I,II}$ to leading order}

Far from any avoided crossing for the dressed two-atom states, the contribution of $\hat{V}_\mathrm{dd}$ to $\Delta E_{I,II}$ is treated perturbatively to first order. Then, the contributions of $\hat{h}_0$, $\hat{v}_Z$, and $\hat{v}_S$ cancel out in  $\delta E_{n_1\mathrm{C},n_2\mathrm{C}}$, and $\hat{V}_\mathrm{dd}$ is neglected if $n_1=n_2$. Therefore, $\Delta E_{I}=\Delta E_{II}\approx-V_{\mathrm{dd}}^{(51)}$, where the matrix element $V_{\mathrm{dd}}^{(n)} = \bra{n\mathrm{C},(n+1)\mathrm{C}} \hat{V}_\mathrm{dd}\ket{(n+1)\mathrm{C},n\mathrm{C}}$, evaluated for zero external fields. We calculate it exactly  using the known matrix elements for the hydrogen atom \cite[\S 36, \S 106, \S 107]{landau3:BH1997}:
\begin{align} \label{eq:VDD0}
\frac{V_{\mathrm{dd}}^{(n)}}{E_\mathrm{H}} &= \frac{3\cos^2\theta-1}{(d/a_0)^3} \frac{n^4}{4} 
  \left(1+\frac{1}{n}\right)\left[1-\frac{1}{(2n+1)^2}\right]^{2n+3} \\
  &= \frac{3\cos^2\theta-1}{(d/a_0)^3} \frac{n^4}{4} \left[1 + \frac{1}{2n} + O(1/n^2) \right] \ ,   
\end{align}
with $\theta$ being the angle between the quantization and interatomic axes. Hence, the interaction between the two atoms is  well represented by
\begin{align}
V_{\mathrm{dd}}^{(n)} &= C_{3,\mathrm{H}}^{(n)} \frac{3\cos^2\theta-1}{d^3} \, , \quad
  C_{3,\mathrm{H}}^{(n)} = \mathrm{Ry} \frac{a_0^3 n^4}{2}.
\label{eq:C3H}
\end{align}
This formula corresponds to the dashed lines in Figure~\ref{fig:interactions}.

\subsection{\label{sec:numerical}
  Numerical approach with multiple
  basis states}
  
  \begin{figure}[t!]
  \centering
  \includegraphics[width=.99\linewidth]
  {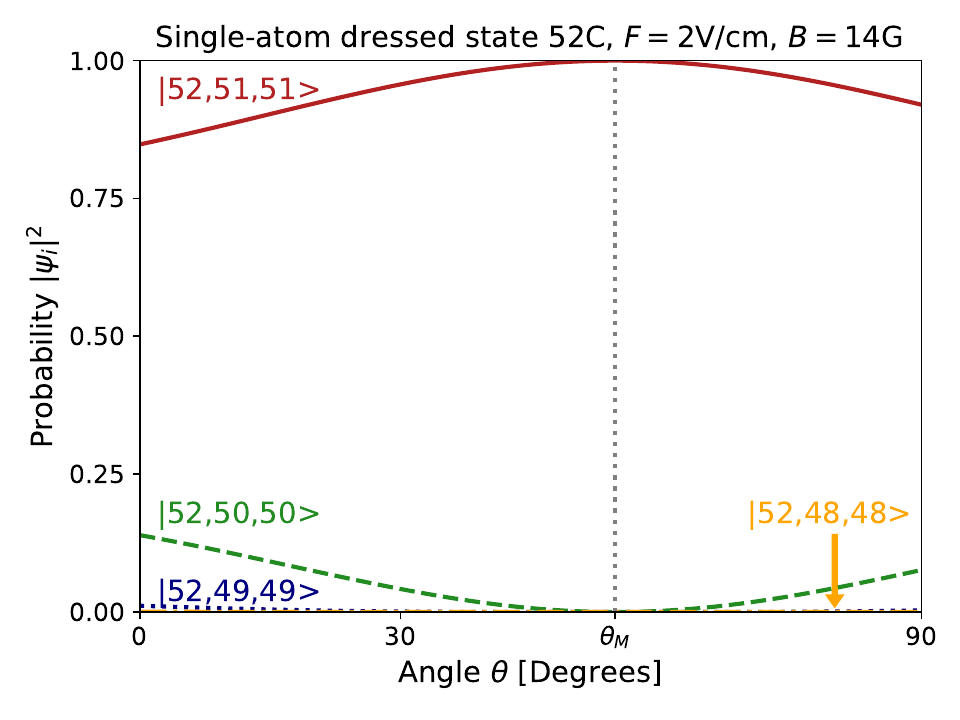}
  \caption{ \label{fig:overlaps}
    Squared overlaps $|\braket{n,\ell,m|52\mathrm{C}}|^2$ of the dressed state $\kcirc{52}$ with the four bare states $\ket{n,\ell,m}$ whose contributions are maximal, as a function of the angle $\theta$ between $\vv{F}$ and $\vv{n}$, for the field magnitudes $F=2\,\mathrm{V/cm}$ and $B=\SI{14}{\gauss}$ used in the experiment and for $d=\SI{13}{\micro\meter}$. The probability $|\braket{52,48,48|52\mathrm{C}}|^2$ (orange curve) always remains lower than $0.001$.
    }
\end{figure}

In this second approach, we account for dressed-state effects. We confirm  the   absence  of   avoided  crossings  for   the  considered parameters and we calculate $\Delta E_{I,II}$ numerically. We calculate  the energies $E_{n_1\mathrm{C},n_2\mathrm{C}}$, with $n_1\leq n_2$, entering  Eq.~(\ref{eq:DeltaEI_DeltaEII}), taking into account the angle between the magnetic field and the interatomic axis, equal to $\theta_0 = \theta_m$.  We restrict the two-atom   Hamiltonian   $(\hat{h}^{(1)} + \hat{h}^{(2)} + \hat{V}_\mathrm{dd})$  to
suitable  subspaces of  two-atom states. To construct the separate subspaces, whose dimensions are collected in Table \ref{table:basissizes}, used for the calculation of the energy $E_{n_1\mathrm{C},n_2\mathrm{C}}$, we first select a basis of single-atom states.  For all considered values of the  external fields, the circular state $\kcirc{n}$, dressed by $\hat{v}_Z$ and $\hat{v}_S$, is well represented by a  linear superposition of  the bare states  $\ket{n,\ell,m=\ell}$ with $n-4\leq  \ell\leq  n-1$ (see Fig.~\ref{fig:overlaps}). Thus, we start  with the angular momenta $\ell,m=\ell$ with  $n_1-4 \leq \ell \leq  n_2-1$,  and further  include  all  angular momenta  $\ell',m'_{l}$  coupled  to them  through  $\hat{V}_\mathrm{dd}$ up  to  the first  (smaller basis) or  second (larger basis) order.   In all cases, we  retain the principal quantum numbers $n_1-4\leq n' \leq n_2+5$ (smaller basis) or $n_1-5\leq  n'  \leq n_2+5$  (larger  basis).   For the  larger  basis related  to  $E_{51\mathrm{C},52\mathrm{C}}$,  we  also  include  states  with  $n'=58$. Turning to two-atom states, we account for all symmetric states built from the above single-particle  basis whose principal quantum numbers $n'_1$ and  $n'_2$ satisfy $|n_1'+n_2'-n_1-n_2|\leq 2$.   The reported values  of  $E_{n_1\mathrm{C},n_2\mathrm{C}}$ calculated  using  the  larger basis  were obtained within a few hours on a recent desktop computer.    

\begin{table}[b]
  \begin{center}
    \begin{tabular}{|l|c|c|}
      \hline
      & {smaller basis} & {larger basis} \\
      \hline
      $E_{52\mathrm{C},52\mathrm{C}}$ & $1470$ & $4700$ \\
      \hline
      $E_{51\mathrm{C}, 52\mathrm{C}}$ & $2193$ & $6820$ \\
      \hline
      $E_{51\mathrm{C}, 51\mathrm{C}}$ & $1470$ & $4700$ \\
      \hline
    \end{tabular}
    \caption{\label{table:basissizes} Numbers of basis states used in the numerical calculations described in Sec.~\ref{sec:numerical}.}
  \end{center}
\end{table}

For  all reported  numerical calculations, we plot in Fig.~\ref{fig:scaledEnergy}, for three different values of $d$, the calculated energy $\Delta E_I$ rescaled by $C_{3,\mathrm{H}}^{(51)}/d^3$, where $C_{3,\mathrm{H}}^{(51)}$ stems from the analytical prediction of~\eqref{eq:C3H}. We compare the  results obtained  with  a smaller  number of  states, chosen to  account for first-order couplings  due to $\hat{V}_\mathrm{dd}$, and a larger one, chosen  to account for second-order couplings.  Numerical  results  obtained with  the  smaller  and larger  subspaces coincide,  and they  are in  excellent agreement  with the  analytical prediction. We also plot in Fig.~\ref{fig:scaledEnergy} the experiment results from Fig.~\ref{fig:interactions}(a) rescaled by $C_{3,\mathrm{H}}^{(51)}/\tilde{d}^3$, taking into account the experimentally rescaled distance. The analytical and numerical predictions are in very good agreement with the experimental data.

\bibliography{Mehaignerie2024} 

\end{document}